\documentclass{jaa}

\usepackage{graphicx}
\usepackage{graphics}
\usepackage{color}
\usepackage{colortbl}
\usepackage{rotating}
\usepackage{amsmath,amssymb,mathrsfs}

\newcommand{\crsf}{cyclotron resonance scattering feature \,}
\newcommand{\rxte}{\textit{RXTE}}
\newcommand{\suz}{\textit{Suzaku }}
\newcommand{\beppo}{\textit{BeppoSAX}}
\newcommand{\integ}{\textit{Integral}}
\newcommand{\nustar}{\textit{NuSTAR}}

\newcommand{\acp}{accretion powered pulsars }

\newcommand{\obss}{observations }
\usepackage{aas_macros}
\usepackage{natbib}

\begin{document}

\title{Cyclotron lines: from magnetic field strength estimators to geometry tracers in neutron stars}


\author{Chandreyee Maitra\textsuperscript{1}}
\affilOne{\textsuperscript{1}Max-Planck-Institut f{\"ur} extraterrestrische Physik, Giessenbachstra{\ss}e, 85748 Garching, Germany}


\twocolumn[{

\maketitle

\corres{cmaitra@mpe.mpg.de}

\msinfo{1 January 2015}{1 January 2015}{1 January 2015}

\begin{abstract}
With exactly forty years since the discovery of the first cyclotron line in Her X-1, there have been remarkable advancements in the field related to study of the physics of accreting neutron stars -- cyclotron lines have been
a major torchbearer in this regard, from being the only direct estimator of the magnetic field strength, a tracer of accretion geometry and an indicator of the emission beam in these systems.
 The main flurry of activities have centred around studying the harmonic separations, luminosity dependence, pulse phase dependence and more recently the 
shapes of the line and the trend for long term evolution in the line energy. This article visits the important results related to cyclotron lines since its discovery and reviews their significance. An emphasis is laid
on pulse phase resolved spectroscopy and the important clues a joint timing and spectral study in this context can provide, to build a complete picture for the physics of accretion and hence X-ray emission in accreting neutron stars.
\end{abstract}


}]


\doinum{12.3456/s78910-011-012-3}
\artcitid{\#\#\#\#}
\volnum{123}
\year{2016}
\pgrange{23--25}
\setcounter{page}{23}
\lp{25}

\section{Introduction}
At the final evolutionary stage of massive stars exceeding the Chandrasekhar mass, matter is compressed to its extreme limit. This results in a compact stellar object, the `neutron star' (NS from now) where the tremendous gravitational pull is halted by repulsive interactions of the degeneracy pressure of the nucleons. A substantial fraction of the known neutron stars reside in X-ray binaries, providing an ideal site to study these objects. Neutron star binary systems; aka accretion powered pulsars (ACPs),
 accrete matter from the companion and emit pulsed radiation at X-ray wavelengths. Accretion powered X-ray pulsars are some of the most powerful sources of X-ray radiation in our Galaxy. Their luminosity lie within $10^{33}-10^{35}  \mathrm{erg}\, \mathrm{s}^{-1}$ during quiescence, and can rise up to $10^{38}  \mathrm{erg}\, \mathrm{s}^{-1}$ during the active state, and they radiate this energy in a broad energy range of $\sim$ 0.1-100 keV.  
 This broadband nature of the energy spectrum originates from multiple and complicated processes which occur near the NS surface, where the infalling
accreted matter is brought to rest and X-rays are emitted by conversion of the gravitational potential
energy to heat energy.

The strong magnetic fields of \acp of the order of $\sim$ $10^{11}-10^{13}$ G have a crucial contribution
 in the formation of the X-ray spectrum of these sources. At the magnetic poles of the NS, the soft photons produced in the thermal mound are Comptonized
in the accretion shock emitting X-ray and Gamma ray photons. Due to the effect of a non-spherical emission zone and
 the scattering cross sections of the photons which are altered 
in the presence of strong magnetic fields (i.e. acquire different dependencies depending whether
they escape parallel or perpendicular to the magnetic field vector \citep{becker1998}), different beaming patterns of radiation, `Fan' or `Pencil' are produced depending on the
rate of mass accretion on the NS and hence the luminosity produced. An unique characteristic
which arises due to the modification of the cross sections by the magnetic field is the formation of the \crsf(henceforth CRSF) which is the focal point
of discussion in this article. Their formation can be described briefly as follows: The
 photons  in the accretion column also undergo scattering with the electrons in the mildly 
relativistic plasma in the accretion column ($\sim$ 0.4c). 
In presence of magnetic fields these electrons exhibit a helical trajectory. The Larmor radius and the corresponding Larmor frequency is given by 
\begin{equation}
r_{b}=\frac{m_{e}v_{\perp}}{eB}
\end{equation}
and
\begin{equation}
\omega_{b}=\frac{eB}{mc}
\end{equation}

where 
 $m_{e}$ denotes the electron mass, $v_{\perp}$ the perpendicular component of velocity of the electron, $e$, 
 the charge of an electron and $B$, the magnetic field strength of the NS. 
 
 At high magnetic fields of the order of the critical magnetic field strength (where cyclotron energy equals the
 electron rest mass energy),
$B_{crit}=\frac{m^{2}c^{3}}{e\hbar}=44 \times 10^{12}$ G. In this condition Larmor radius $\sim$
De Broglie radius
of the electron in the plasma.
 A relativistic quantum mechanical treatment becomes necessary as the electron's perpendicular momenta are
quantized into discrete Landau levels. This is given as
\begin{equation}
 \frac{p_{\perp}}{mec}=n\frac{B}{B_{crit}} 
 \end{equation}
 The scattering cross section is
therefore resonant at the energy separation of the Landau levels. Due to the very high cross sections at these
resonances, and the fact that the Landau levels are thermally broadened, a photon with its energy close to the Landau level
separation may not escape the line forming region, and therefore be consequently observed as an absorption feature in the energy
spectrum. The energy is given by 

\begin{equation}
E_{n}=(m_{e}c^{2})\frac{\sqrt{1+2n\frac{B}{B_{crit}}\sin^{2}\theta}-1}{\sin^{2}\theta}\times \frac{1}{1+z}
\end{equation}

 $m_{e}$ is the electron mass, $c$ the speed of light, $\theta$ the direction between the incident photon
and the magnetic field vector, $z$ gravitational
radius corresponding to the line forming region ($z=\frac{1}{\sqrt{1-\frac{2GM}{Rc^{2}}}}-1$).

$n=1$ corresponds to the cyclotron fundamental and the corresponding 
harmonics are given by n=2,3,4 etc. The above expression also indicates that the line energy is angular
dependent. The angular dependence of the CRSF provides further handle on the geometry of the accretion region
as will be discussed later. 

\begin{figure*}
	\centering
        \includegraphics[width = 8cm, height = 8cm,keepaspectratio] {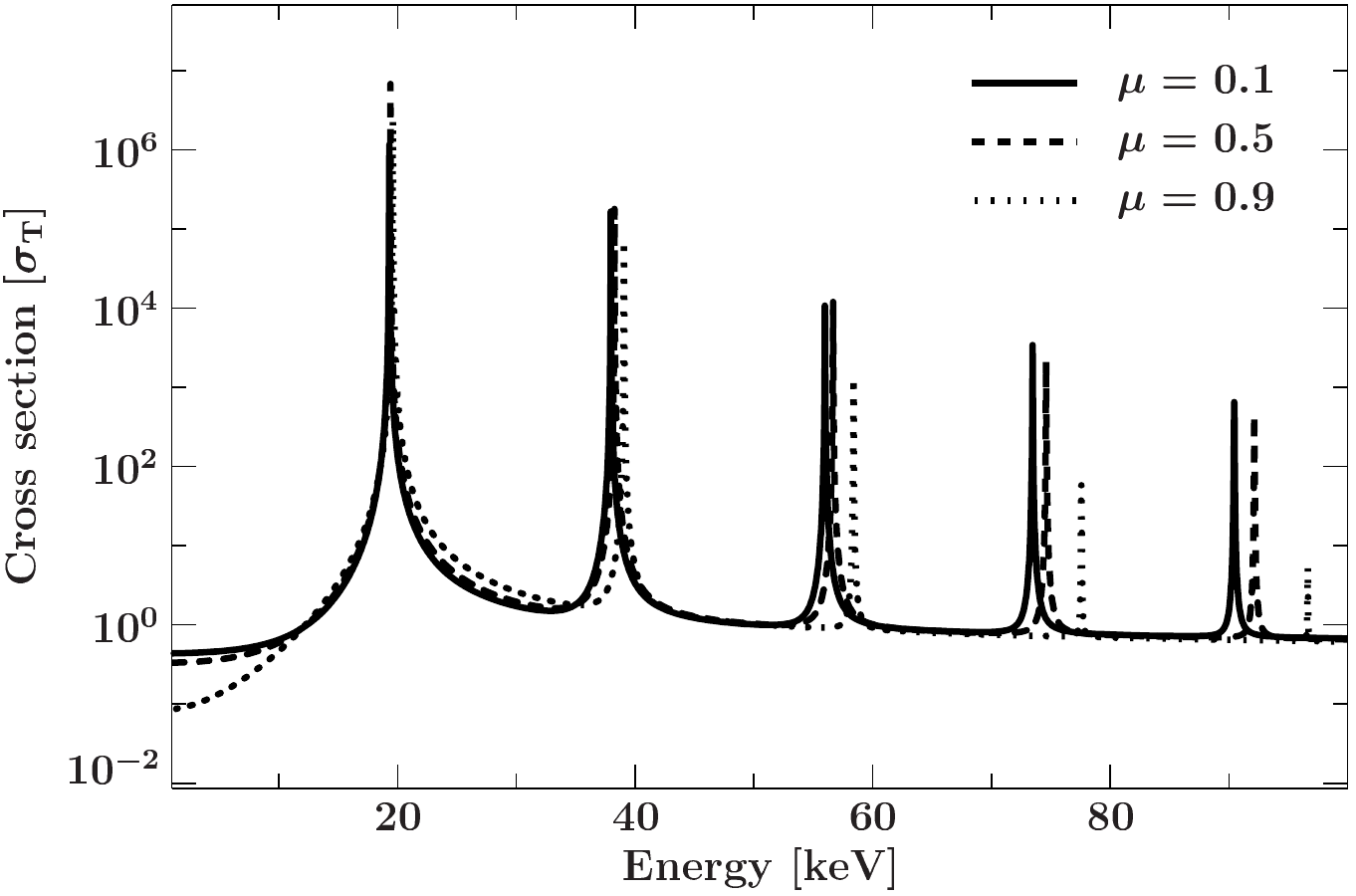}
	\caption{ \small Scattering cross section of CRSF
    for exciting an electron in initial Landau level $n_i=0$ (fundamental).  The solid, dashed, and dotted lines show cross sections for different viewing angles
    $\mu = \cos\vartheta = 0.1, 0.5, 0.9$. Figure adapted from \cite{2017A&A...597A...3S} (Fig. 1)}\label{sc}
\end{figure*}

The presence of the CRSF was first theoretically predicted by \cite{gnedin1974},
even before it was discovered observationally. Subsequently, the formation of CRSFs have been simulated by three main methods: a. Monte Carlo \cite{araya1999,araya2000,schonherr2007} 
b. Feautrier methods \citep{meszaros1985a,nishimura2008} and c. analytically \citep{1993ApJ...414..815W}. The most recent results are Monte Carlo simulation
of the CRSFs  to generate synthetic cyclotron lines based on interpolation tables of photon mean free path for typical magnetic field strengths and electron temperatures relevant
to these systems \cite{2017A&A...597A...3S,2017arXiv170107669S}.  This is a step forward to calculate the expected CRSF for any complex X-ray pulsar geometry given its magnetic field strength and the viewing angles.
Figure \ref{sc} shows the resonant cross sections at different viewing angles as
computed for a magnetic field strength of $1.694\times10^{12}$ G \citep{2017A&A...597A...3S}. 

This article aims at reviewing the importance of CRSFs particularly in the context of studying the magnetic field and geometry of accretion powered
pulsars. Section 2 describes the census of CRSFs and reviews the key results related to the study of CRSFs till now. Section 3 reviews the results on pulse phase
resolved spectroscopy and studies of changes in the pulse shape near the CRSF energies. Section 4 presents a case study of correlated timing and spectral study to study the accretion geometry in an ACP.
Section 5 describes the conclusions.

\section{CRSFs: key results}
Due to the effect of gravitational redshift near the NS, the CRSF are actually observed at a centroid energy of $E_{obs}=E(1+z)$.
For typical NS parameters, z  is $\sim 0.3$. \\
Cyclotron line was first discovered in the \acp Her X-1 \citep{trumper1977}. Initially thought to be in emission at $\sim 53$ keV, it
was later inferred to be in absorption at $40$ keV based on theoretical arguments \citep{nagel1981}.  This provided the first direct measurement of the
 magnetic field strength of a NS. A pioneer
in the observational studies of \crsf has been the satellite \emph{Ginga} \citep{mihara1995}, and CRSFs in many sources
have been discovered by this space mission. Subsequently
with the satellites \rxte, \beppo, \integ, and specially
\suz and more recently \nustar~ providing high quality spectral data over a broad energy range, there have been discoveries of CRSFs in new sources
and many followup observations have been carried out, specially during the outbursts and bright states of the sources. This enabled detailed studies on CRSFs and a significant advancement in this field. 

CRSFs  are usually detected as absorption lines in the continuum spectrum, and are modelled phenomenologically with Gaussian or pseudo-Lorentzian profiles. 
This provides information on the line centroid energy and the equivalent width of the feature.  A careful modeling of the broadband continuum spectrum is required  for obtaining consistent and physically 
 reliable parameters. The continuum model should not introduce any artificial  residual like structure in the spectrum which can be mistakenly modelled as a part of the
 cyclotron line. 
MC simulations by \cite{2017A&A...597A...3S} have also made available a physical model which allows to compare the observed feature with a synthetic cyclotron line for an arbitrarily complex cylindrically symmetric geometry, and given physical parameters
of the accretion column and the magnetic field strength of the NS. 

\subsection{CRSFs and harmonics: current census}

The current census is  more than 30 CRSF sources discovered with magnetic field strengths measured ranging from $\sim$ 1-7 $10^{12}$ G. Table \ref{crsf-c} gives a list of confirmed CRSF sources \footnote{http://www.sternwarte.uni-erlangen.de/wiki/doku.php?id=cyclo:start}.

Harmonics of the CRSFs are also seen in several sources:  for example up to the fifth harmonic in 4U 0115+63 \citep{santangelo2000,heindl2004}, two harmonics in V 0332+53  \citep{potts2005},
and a single harmonic in several sources like Vela X-1 \citep{kreykenbohm2002},  A 0535+26 \citep{caballero2007} and 4U 1538-52 \citep{2009A&A...508..395R}.   \cite{Iyer2015} has reported the presence of two independent sets of fundamental CRSF in 4U 0115-63, possibly originating from spatially distinct emitting regions.
An-harmonic spacing between the lines have been seen in some sources where the ratio
of the centroid energy of the harmonics deviate from the expected value of 2. Deviations $< 2$ can be explained by relativistic corrections as in 4U 0115+63 \citep{santangelo2000} and $> 2$ by superposition of
many lines as in Vela X-1 \citep{kreykenbohm2002} and A 0535+26 \citep{caballero2007}. Recently \cite{2015MNRAS.453L..21J} discovered the first harmonic of the CRSF in Cep X-4 at a ratio of 1.7 which they interpret
as a result of different heights of the two line forming regions, or a larger viewing angle of the system.
The fundamental profile usually appears shallower than its corresponding
higher order harmonic. 

\subsection{Indication for complex shape of CRSF}
Complex CRSF shapes have been predicted by theory including the presence of  line `wings' in the fundamental \citep{schonherr2007,2017A&A...597A...3S,2017arXiv170107669S}.
A complex CRSF fundamental (deviating from a standard Gaussian or Lorentzian profile while fitting the absorption feature)
was reported in V 0332+53 \citep{nakajima2010} using~\rxte~observations. Later however \cite{2017MNRAS.466.2143D} found no strong evidence for a more complex line profile in ~\nustar~ data,
given the existing uncertainties in the modelling of the broad-band continuum of X-ray pulsars. 
A complex shape of the CRSF was also reported in Cep X-4 \citep{cepx4} using a~ \nustar ~observation. Recently \cite{2017MNRAS.470..713M} found signatures of a non Gaussian CRSF 
in low luminosity pulsar X~Persei from a~\suz observation of the source. Complex shapes are expected to be more prominent at lower luminosities. \cite{mukherjee2011} predicted 
complicated line shapes in the spectra of low luminosity pulsars due to the distortion of the dipolar magnetic field near the base of the accretion mound due to 
the effect of local instabilities. \cite{Mushtukov2015} also predicted a similar effect for low luminosity sources  due to the variation of the Doppler boost with 
changing luminosity. The observed CRSF in the low luminosity source X-Persei is consistent with these predictions. More such discoveries and comparison of the line profile with physical models will enable to bridge the gap between predictions from theory and observational results. 
\subsection{Variations with luminosity}
Variations of the CRSF parameters have been observed with luminosity on pulse to pulse timescales as well as on longer timescales. Especially the line centroid energy exhibits either a correlation or anti-correlation with luminosity, with only a few sources exhibiting no change in the centroid energy with luminosity. The observed bimodality is widely believed to be due to two different accretion regimes depending on the `critical luminosity' of the sources:  $L_{\mathrm{crt}}$  which divides two regimes of accretion into the super and sub-critical regime \citep{basko-sunyaev1976,becker2012}. The expression is given by: 

\begin{eqnarray}
	L_{\rm crt} &=& 1.49 \times 10^{37}{\rm erg\,s}^{-1} \left( \frac{\Lambda}{0.1} \right)^{-7/5} w^{-28/15} \nonumber \\
	&&\times \left( \frac{M}{1.4{\rm\,M}_{\odot}} \right)^{29/30} \left( \frac{R}{10{\rm\,km}} \right)^{1/10} \left( \frac{B_{\rm surf}}{10^{12}{\rm\,G}} \right)^{16/15}
	\label{eqn:lcrit}
\end{eqnarray}
Here, $M$, $R$, and $B$ are, the mass, radius, and surface
magnetic field strength of the NS, $w = 1$ characterizes the shape of the
photon spectrum inside the column (with mean photon
energy of $\bar{E} = wkT_{\rm eff}$), $\Lambda$ characterizes the mode of
accretion. $\Lambda= 0.1$ approximates the case of disk accretion, whereas $\Lambda= 1.0$ is more appropriate for wind accretors. $L_{\rm crt}$ is therefore higher for higher magnetic field strengths of the NS, and is higher for wind accretors compared to disk accretors. 
 Assuming accretion via the disk and a typical magnetic field strength of $B =10^{12}$ G, $L_{\mathrm{crt}}$ $\sim 10^{37}$ $erg s^{-1}$.  At L $>$ $L_{\rm crt}$ {\it i.e.} the supercritical regime, an accretion column is formed and the radiative shock  decelerates the infalling matter forming a `Fan' like beam pattern. Therefore height of the line forming region increases with increasing mass accretion rate, sampling a net lower magnetic field strength. This explains the negative correlation between the cyclotron line energy and the luminosity.   At L $<$ $L_{\rm crt}$ {\it i.e.} subcritical regime, Coulomb interactions stop the infalling matter forming a `Pencil' like beam pattern, and the height of the line forming region decreases with increasing accretion rate. This explains the positive correlation observed. The above model requires a gradient in the magnetic field strength
in the line forming region to explain the observed variations with luminosity. Alternate theories to explain the correlation of centroid energy with luminosity include \cite{2013ApJ...777..115P}
where the observed anti-correlation with luminosity are explained by variations of the NS area illuminated by a growing accretion column. This model requires the line to be formed when the radiation is reflected from the NS
surface, and can explain a smaller variation with luminosity than predicted by the former model. \cite{2014ApJ...781...30N} proposed that changes in the emission pattern towards `Fan' like beam at higher luminosity affect the position of the CRSF via the Doppler effect explaining the negative correlation with luminosity. \cite{Mushtukov2015} further incorporated the changes in velocity profile in the line-forming region due the radiation pressure force to explain the observed positive correlation of CRSFs with luminosity also via the Doppler effect.

 Sources which exhibit a positive correlation of the CRSF centroid energy with luminosity are Her X-1 \citep{staubert2007}, GX 304-1 \citep{2012A&A...542L..28K}, and Swift J1626.6-5156 \citep{2013ApJ...762...61D}, while sources exhibiting and opposite trend are 4U 0115+6415 \citep{nakajima2006}, V 0332+53 \citep{tsygankov2006} and SMC X-2 \citep{2016MNRAS.461L..97J}. The relationship can also be more complex in nature. The centroid energy of the CRSF in
A 0535+26 shows no correlation with luminosity \citep{2013ApJ...764L..23C, 2006ApJ...648L.139T}.  \cite{klochkov2011} however reported a positive correlation with luminosity for this source in certain pulse phase bins. The source may also have a positive correlation in phase-averaged spectra at higher luminosities \citep{2015ApJ...806..193S}. \cite{2014ApJ...780..133F} found that the energy of the first harmonic of the CRSF in Vela X-1 was positively correlated with luminosity, while the behaviour of the fundamental was more difficult to ascertain.  \cite{2017MNRAS.466.2143D} found signatures of transition from super to sub-critical accretion for the first time in a source in V 0332+53 though its departure
from the expected negative correlation of the CRSF centroid with energy.

There is one fact which is crucial to keep in mind when estimating the correlation of the centroid energy of CRSF with luminosity. The CRSF centroid energy estimated is dependent on the continuum spectrum and the correlation has to be computed with the same continuum spectral model, as well as the same line shape (Gaussian vs. Lorentzian). \cite{muller2013} discusses this in detail and report a different dependence of the CRSF energy with luminosity in the case of 4U 0115+63 using two different continuum models.
\subsection{long term secular evolution of the CRSF centroid energy}
An aspect which has instilled a lot of interest in the community is the evidence of a long term change in the centroid energy of the CRSF. Although this has been seen in a few sources until now, the potential of 
discovering this trend in other sources is very promising, albeit a careful and systematic scrutinisation of the long term behaviour of sources. Her X-1 showed an evidence of a true long-term decrease in the centroid energy of the CRSF in the pulse phase averaged spectra
 from 1996 to 2012 \citep{staubert2014}. The decay can be modelled as either by a linear decay, or by a slow decay until 2006 followed by a more abrupt decrease after that. The reason has been speculated to be 
 connected to a geometric displacement of the line forming region or to a physical change in the magnetic field configuration at the polar cap due to the process of continued accretion.
 On the contrary, it was observed in 4U 1538-52 \citep{2016MNRAS.458.2745H} that the CRSF centroid energy increased between the \rxte measurements of 1996 to 2004 and the 2012 \suz 
 measurement. The increase of 5\% in the CRSF energy can be attributed
 to a true reconfiguration of the magnetic field at the base of the accretion column or the accretion mound, therefore sampling a stronger average field strength \citep{mukherjee2011}. V 0332+53 also exhibited a sudden drop in its CRSF energy during a giant outburst in 2015 \citep{2016MNRAS.460L..99C}. However it was noticed that a year later, the line energy increased again reaching the pre-outburst values. 
 This behaviour was attributed to be likely caused by a change of the emission region geometry rather than an accretion-induced decay of the neutron stars magnetic field \citep{2017MNRAS.466.2143D}.

\begin{table*}
	\caption{Cyclotron Resonance Scattering Feature Table in increasing order of line centroid energy. P/T: persistent, transient. Only confirmed sources with high detection significance of the CRSF are tabulated here.}
	\label{crsf-c}
	\centering
	\scriptsize
        \begin{tabular}{c  c  c   c }
        \hline \hline
	Source name & Energy (keV) & Companion & Source Type (P/T) \\
	\hline 
	XMMU J054134.7-682550 & 9 & Be? & T\\
	Swift J1822.3-1606 & 10--15 ? & B1V  & T \\
	KS 1947+300 & 14 & B0V  & T\\
	Swift J1626.6-5156 & 10 & Be & T \\
	4U 0115+63 & 14, 24, 36, 48, 62 & BO.5 Ve & T \\
	IGR 17544-2619 & 17 & B1.0 & P (SFXT) \\
	4U 1907+09 & 18, 38 & B2 III-IV & P \\
	4U 1538-52 & 22, 47 & B0I & P \\
	2S 1553-542 & 23.5 & Be? & T\\
         Vela X-1 & 24, 51 &  B0.5Ib & P \\ 
	V 0332+53 & 27, 51, 74 & O8.5 Ve & T \\
	SMC X-2 & 27 & B1.5 Ve & T \\
	Cep X-4 & 28 & B1.5 Ve & T \\
	IGR J16393-4643 & 29 & BIV-V  &  P\\
	Cen X-3 & 30 & O6.5II & P \\
	IGR 16393-4643 & 29 & B0 V & P\\
	RX J0520.5-6932 & 31.5  & O8Ve & T\\
	RX J0440.9+4431 & 32 & Be & T \\
	MXB 0656-072 & 33 & O9.7 Ve & T \\
	XTE J1946+274 & 36 & B0-1 V-IVe & T \\
	4U 1626-67 & 37 & WD? & P \\
	GX 301-2 & 37 & B1.2Ia & P \\
	Her X-1 & 41 & A9-B & P \\
	MAXI J1409-619 & 44, 73, 128 & B0 & T\\
	1A 0535+26 & 45, 100 & O9.7 IIe & T \\
	1A 1118-61 & 50, 110 ? & O9.5IV-Ve & T \\
	GX 304-1 & 54 & B2 Ve & P \\
	GRO J1008-57 & 76 & B1-B2 & T \\
	
\hline
	\end{tabular}
	\end{table*}

\section{Pulse phase resolved spectroscopy: a geometry tracer}
Inherently, the angular dependence of the CRSF cross section is expected to result in variations of a few
$\%$ in the cyclotron parameters with the viewing angle of the NS: the pulse phase (as evident from
Equation 4). In addition, as the CRSF depends on the physical parameters of the region like the
plasma temperature, the optical depth and the geometry of the region, projections of different parts of the accretion
column having different physical
properties, can also result in variations of the CRSF parameters with pulse phase.
In \cite{Mushtukov2015}, changes in the velocity profile in the line forming region with the electron velocity at the base of the accretion channel reaching zero (when the source reaches critical luminosity)
also produce significant change of the CRSF parameters with pulse phase. Pulse phase dependence of the CRSF parameters thus provides crucial clues regarding the line forming region at different viewing angles, providing information on the distribution of plasma temperature and optical depth of the region.  It also provides important clues on the geometry of the line forming region and can also map the magnetic field configuration of the same. 
 
 Pulse phase resolved spectroscopy of the cyclotron parameters have been performed for many sources, for example in Her X-1 \citep{soong1990,enoto2008,klochkov2008},
4U 1538-52 \citep{robba2001}, 4U0115+63 \citep{heindl2000}, Vela X-1 \citep{kreykenbohm1999,kreykenbohm2002,labarbera2003, maitra2013a}, Cen X-3 \citep{suchy2008},  GX 301-2 
\citep{suchy2012}, 1A 1118-61 \citep{suchy2011, 2012MNRAS.420.2307M}, A 0535+26, 4U 1907+09, XTE J1946+274 \citep{2013ApJ...771...96M}, 4U 1626-67 \citep{iwakiri2012} and GX 304-1 \cite{2016MNRAS.457.2749J}. \cite{2014EPJWC..6406008M} also presented a comprehensive results of phase resolved spectroscopy of CRSFs and it's importance.

 A detailed and systematic study of pulse phase resolved spectroscopy was carried out using a sample of relatively bright CRSF sources. For all of these sources, long observations were available with {\it Suzaku}, therefore providing broadband spectral coverage to study the spectrum in detail (Maitra, C Ph.D Thesis, 2013  \footnote{http://hdl.handle.net/2289/6033}). In this work, phase resolved spectra were generated with the phases centered 
around 25 independent bins but at thrice their widths. This resulted in 25 overlapping bins out of which
only 8 were independent. This helped in obtaining a very detailed pattern of variation of the CRSF parameters.  All the individual spectra for the respective sources were fitted with the best fitted
continuum and line models, after checking the consistency between the model parameters.
Table \ref{crsf} summarizes the main results obtained including the \% of variation of the CRSF centroid energy.
The obtained pattern of variation of the CRSF parameters are similar with different continuum models, indicating the robustness of the results.
Some interesting outcomes were as follows: a pulse phase dependence of the ratio of the fundamental and first harmonic in Vela X-1 was detected as seen in Fig. \ref{vela} (indicating the line forming region of the fundamental and the harmonic
varies across the viewing angle, hence across the accretion column); a similar pattern for pulse phase dependence of the CRSF parameters was obtained in 4U 1907+09 at a factor of two difference in luminosity as seen in Fig. \ref{1907}. The two
observations also exhibit very similar pulse profiles indicating that the line forming region is located at roughly the same height from the neutron star surface and have a similar geometry for this range of luminosity.

The results of pulse phase resolved spectroscopy obtained from the sample of sources indicated an overall complex dependence, and typical variations
with phase are 10--30 \% or more. The effect of gravitational light bending due to the strong gravity near the NS surface
\citep{leahy2003} may smear the pulse phase dependence results, with 
a particular viewing angle having contributions from multiple emission regions. This will however only lower
the net \% of variation of the CRSF parameters. The large variations cannot be
explained by the change in viewing angle alone, and if explained by the changing projections at different 
viewing angles of different parts of the accretion column, requires a large gradient in the physical parameters across the emission region.
Similarly, a change in the CRSF centroid energy due to the change in the height of the emission region (and hence
the projected magnetic field strength) at different viewing angles requires an accretion column of height several kilometres. Alternately, the results
can be explained by large polar cap regions (with a wide opening angle of $50-60^{\circ}$), or a more
complex configuration of magnetic field consisting of sharp gradients, and non dipolar components \citep{nishimura2008,mukherjee2011,mukherjee2012}.
This is supported by sharp gradients in the variation of the CRSF parameters observed in some sources, which are indicative of 
 a complicated magnetic field geometry (non-dipolar) or a more complex geometry and beaming pattern.
 It has also been proposed that superposition of cyclotron resonance scattering feature, from different line forming regions can give rise
to a large factor of variation in the CRSF parameters \citep{nishimura2011}. 

 An alternate scenario which can produce a larger variation of the CRSF parameters than predicted theoretically is as follows. The `Fan' or `Pencil' like 
beam patterns depend on the source luminosity (whether   L $>$ $L_{\rm crit}$ or  L $<$ $L_{\rm crt}$), and are ideal approximations of the beaming pattern of the radiation. Both 
may actually coexist in an accretion column, with the base of the column emitting a `Pencil' like beam pattern and the sides emitting `Fan' like beam pattern.  The dominance of the two components are determined by the source luminosity. \cite{kraus2001,leahy2004} interpreted the observed pulse profiles of Cen X-3 and Her X-1 as a combination of
`Pencil' and `Fan' like beaming patterns of radiation emitted from the accretion column on the surface of the NS. In this scenario due to the cosine dependence of the cross section with the viewing angle, the phases indicating narrower and deeper CRSFs at higher centroid energies may refer to the region dominated by `Fan' beaming pattern.  Similarly, the ones indicating more broader and shallower CRSFs
at lower energies to the region may refer to a `Pencil' beaming pattern ( \cite{schonherr2007}
for detailed expressions for the expected angular dependence of the CRSFs with viewing angle). The situation is further complicated by the contribution
from both the magnetic poles of the NS adding up to the net beamed profile. Another point to be considered is the accretion geometry itself. 
In the simplified scenario, the accretion geometry is assumed
to be an emitting  region of slab or cylinder on the surface of the NS \citep{meszaros1985b}.
However, the actual geometry would depend on the 
structure of the region where the accreted matter couples to the magnetic field lines
and further on the nature of the region where the matter gets decelerated on the NS surface.
Ring shaped mounds and partially filled or hollow accretion columns may result if the filling factor and hence
the threading region is smaller. Emission from more of these complicated geometries have been calculated
in \cite{kraus2001}, \cite{leahy2003} and references therein. The simulation code of \cite{2017A&A...597A...3S,2017arXiv170107669S} producing synthetic CRSFs for complex scenarios may be applied 
to test the observational results of phase resolved spectroscopy with simulations for complicated geometries.

\begin{table*}
	\caption{Table summarizing pulse phase resolved variations of CRSF in a sample of sources from Maitra, C Ph.D Thesis (2013)}
	\label{crsf}
	\scriptsize
        \begin{tabular}{l l l l l }
        \hline
	Source name  & $E_{CRSF}$ & $\%$ variation of $E_{CRSF}$ & comments \\
	 & & (keV) & & \\
	\hline
	4U 1907+09 & 18 & 19 & CRSF parameters show similar variation for factor of 2 \\
	& & & difference in $L_{x}$\\
	Vela X-1 &  24, 50 & 27 & Detected variation of the ratio of line energies with phase.\\
	4U 1626-67  & 35 & 12 & --\\
	GX 301-2 &  36 & 13 & Sharp gradients in the CRSF parameters detected. \\
	XTE J1946+274  & 38 &36 & -- \\
	A 0535+26 & 43 & 14 & -- \\
	1A 1118-61 &  47 & 30 & --\\
	\hline 
	
	\hline
	\end{tabular}
	\end{table*}

\subsection{Changes in the pulse profiles near the CRSF}
As the scattering cross sections are significantly altered and increase by a large factor near the CRSF; the corresponding beam patterns and hence the pulse profiles
are also expected to change near the corresponding resonance energies. 
Such probes were attempted  by \cite{tsygankov2006,lutovinov2009,ferrigno2011}. 4U 0115+63 \citep{ferrigno2011} showed significant changes of its
shape near the pulse peak at the CRSF fundamental and its corresponding harmonics. V 0332+53  \citep{tsygankov2006} also showed an 
asymmetrical single peaked structure near its fundamental CRSF energy in contrary to the otherwise double peaked profile.
\cite{lutovinov2009} studied both the intensity and energy dependence of the pulse profiles in the hard X-ray 
range (20-100 \rmfamily{keV}) using \integ \, \obss.
The main outcome  of the work was a general trend of increase of the pulse fractions with energy for all 
the sources, with a local maxima near the CRSF energy for some sources. \cite{2016MNRAS.457.2749J}  also found a phase shift near the CRSF energy in the pulse profile of GX 304-1.
This aspect was studied in detail in Maitra, C (Ph.D thesis, 2013.) The pulse profiles for  many sources near their corresponding CRSF energies were studied,  by creating them centred around the CRSF energy with a width equal 
to the FWHM of the feature measured from the spectrum. 

The results showed that GX 301-2, 1A 1118-61 and Her X-1 exhibited distinct changes in the pulse profile near the corresponding CRSF energies, with the higher and lower energy (w.r.t. CRSF) profiles being similar. XTE J1946+274  and A0535+26  exhibited enhanced beaming and change in shape of the pulse profile near the CRSF band, although the higher energy bands
 did not  have enough statistics to be probed for changes. Cen X-3 showed featureless profile near the CRSF at $\sim$ 28 keV. However it was highly beamed with a shift in the
peak at higher energies (34-50 keV). Sources which seemed to show no changes in the pulse profiles near the CRSF
 fundamental were 4U 1907+09 Vela, X-1 and 4U 1626-67. It is noteworthy that all three have a shallow and narrow fundamental CRSF.  However there is an increase in pulse fraction for 4U 1907+09
at the energy corresponding to the possible first harmonic energy of the CRSF. Fig. \ref{fig211} shows the comparison of pulse profiles for the sources which showed a significant change in the pulse shape/fraction at the CRSF band.
The shape of the pulse profiles near the CRSF energies were investigated theoretically in a detailed numerical study by \cite{2014A&A...564L...8S}. The results predicted strong changes in the pulse profiles of magnetized accretion powered pulsars near the corresponding CRSF
energies. This was attributed to the angular redistribution of X-ray photons due to the effect of cyclotron resonance scattering, combined with relativistic effects. The results also predicted a strong dependence on the accretion geometry.
This therefore remains a potential tool to probe the system's accretion geometry.

 \section{Correlated timing and spectral study as a geometry tracer in ACPs: Vela X-1 a case study}
\begin{figure*}
\centering
\includegraphics[height=12cm, width=9cm]{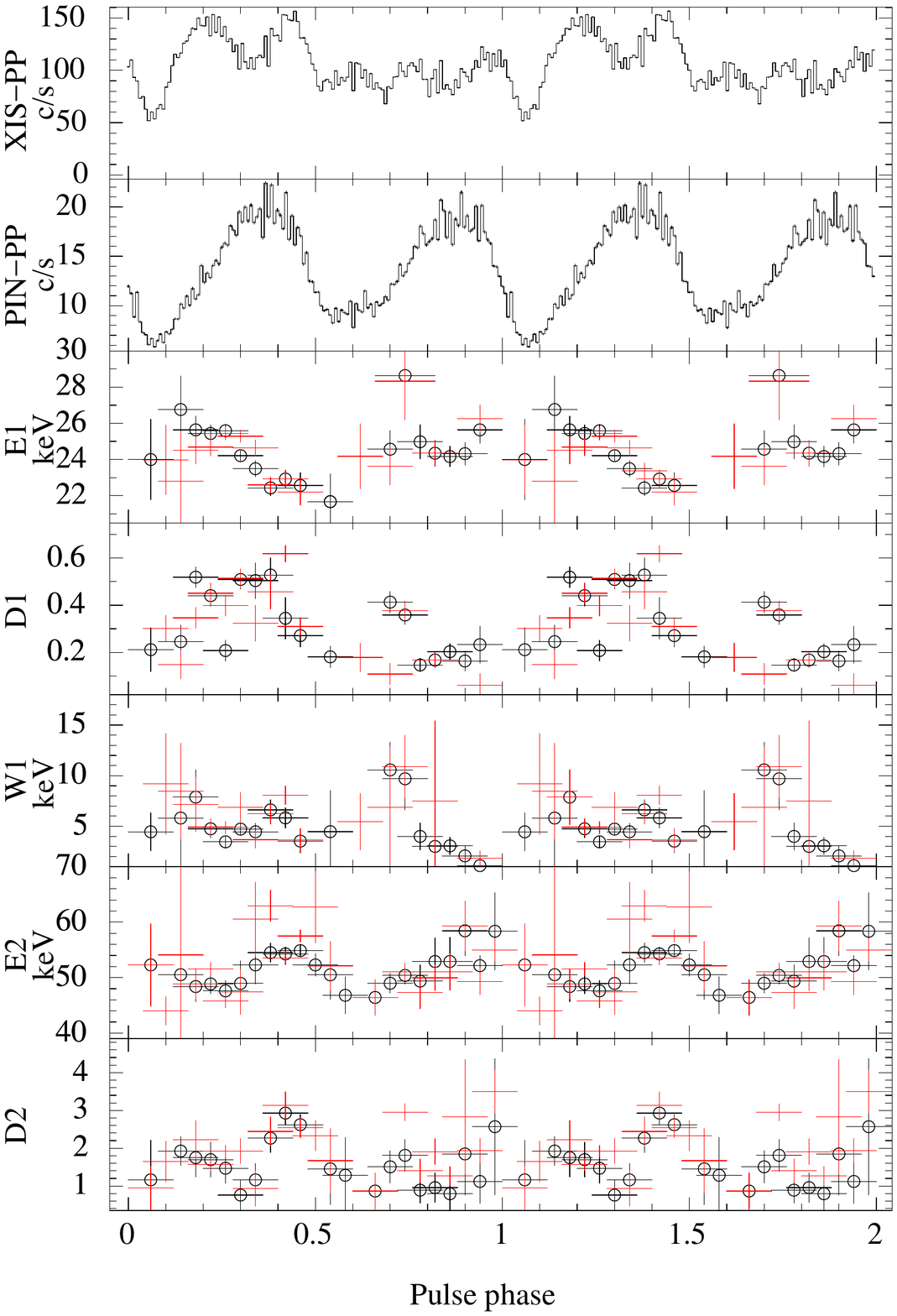}
\includegraphics[scale=0.30]{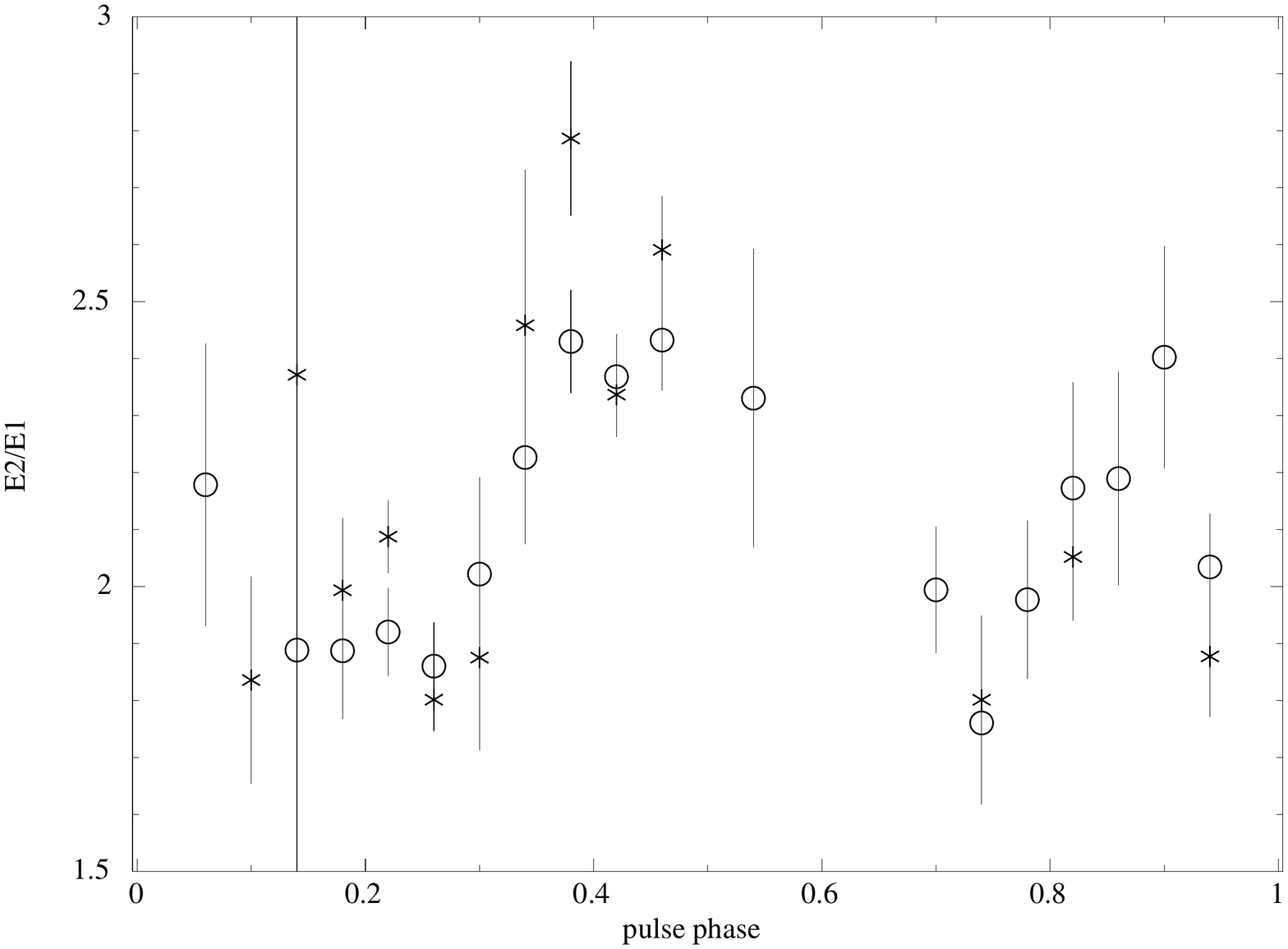}
\caption{Left: Pulse phase resolved spectroscopy of the CRSF parameters in Vela X-1.
The top two panels indicate the normalized intensities of the low energy (0.3-10 keV) and high
energy (10-70 keV) pulse profiles. The black points with 'circle' denote the parameters as obtained with the `Highecut' continuum model. The red points denote the
points as obtained with the 'CompTT' continuum model. E1, D1 and W1 denote the fundamental energy, depth and width respectively. E2 and D2 the energy and depth of the first harmonic. The variations in W2 could not be constrained due to statistical limitations. Only 8 of the 25 bins are independent. Right: Variation of the ratio of the two cyclotron lines E2/E1 as is obtained by fitting the two continuum models.
The variation of the parameter obtained with `Highecut' is denoted by the symbol `circle',  and with `CompTT' is denoted by the symbol `star'. Adapted from \cite{maitra2013a}.}
\label{vela}
\end{figure*}

\begin{figure*}
\begin{center}$
\begin{array}{c c }
\includegraphics[height=10 cm, width=7cm]{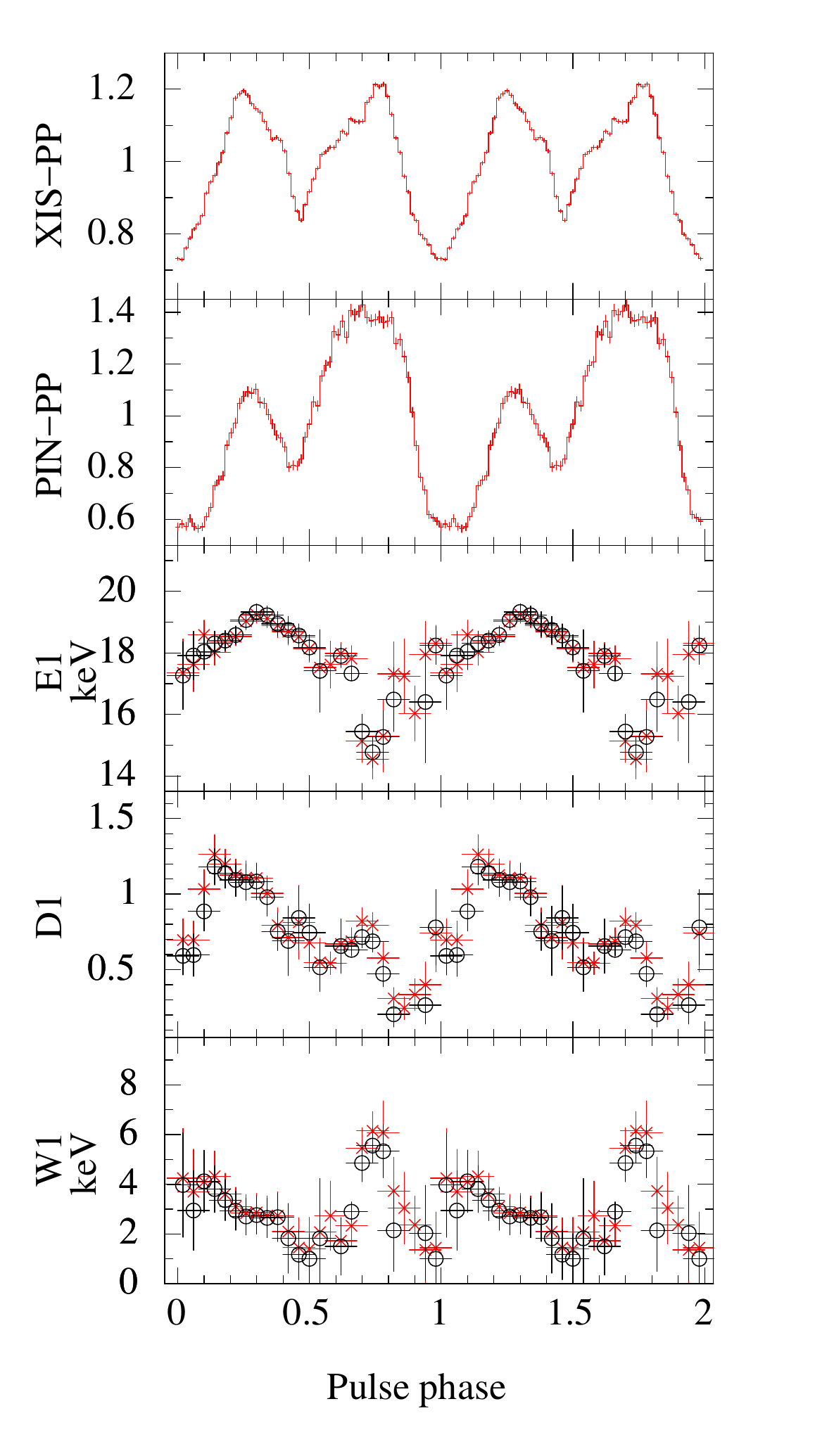} &
\includegraphics[height=10 cm, width=7cm]{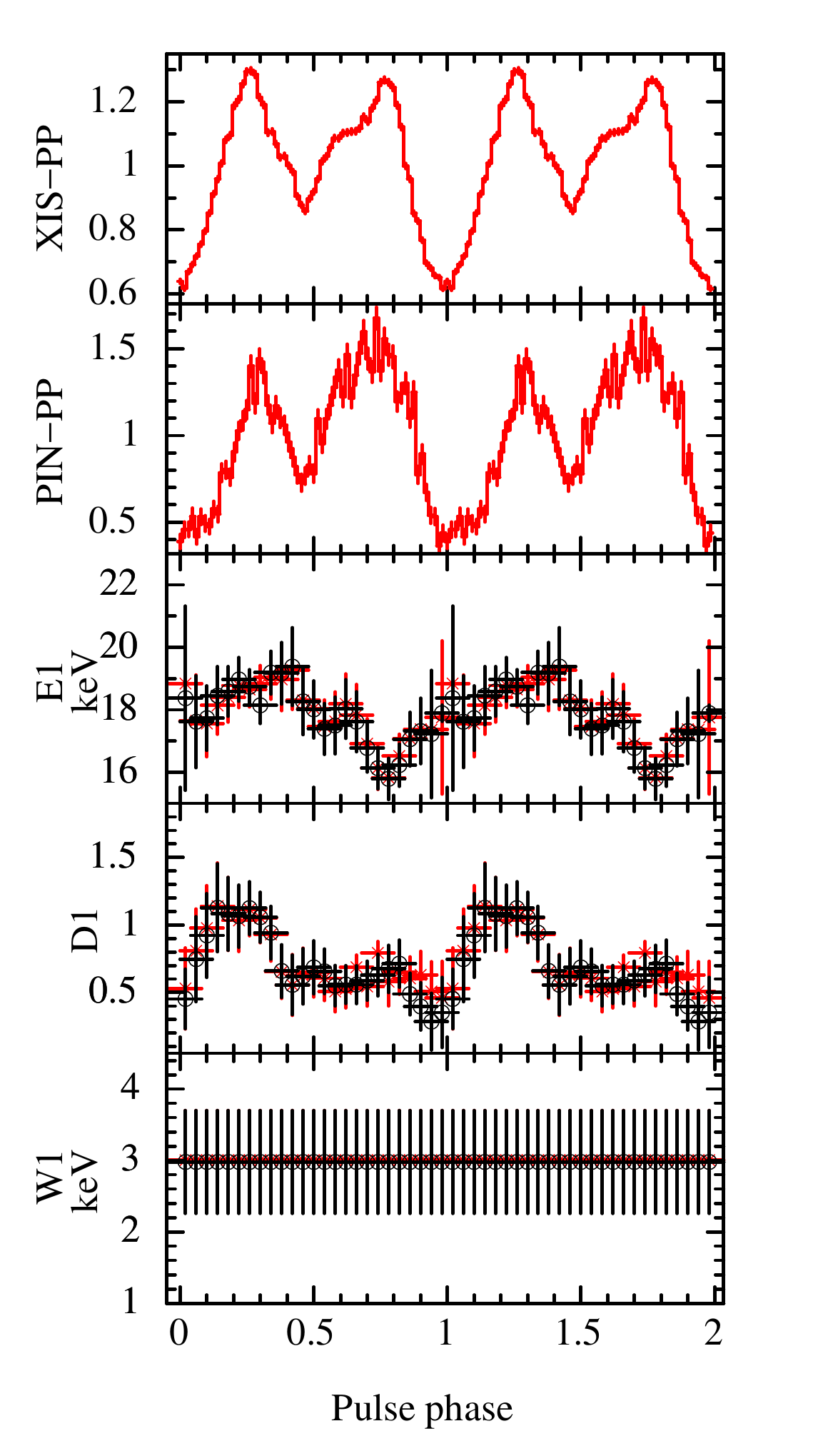}\\
\end{array}$
\end{center}
\caption{Pulse phase resolved spectroscopy for 4U 1907+09 for a factor of 2 difference in luminosity (left figure: higher luminosity). The top two panels indicate the normalized intensities of the low energy (0.3-10 keV) and high
energy (10-70 keV) pulse profiles. 
The black points denote the parameters as obtained with the `NPEX' continuum model. The red points denote the
parameters as obtained with the `CompTT' continuum model.  E1, D1 and W1 denote the CRSF energy, depth and width respectively. Only 8 of the 25 bins are independent. The variation in W1 for the fainter observation 
could not be constrained, and hence fixed to the value obtained from the phase-averaged spectrum. (Maitra, C Ph.D Thesis, 2013)}
\label{1907}
\end{figure*}

\begin{figure*}
\centering
\includegraphics[scale=0.25]{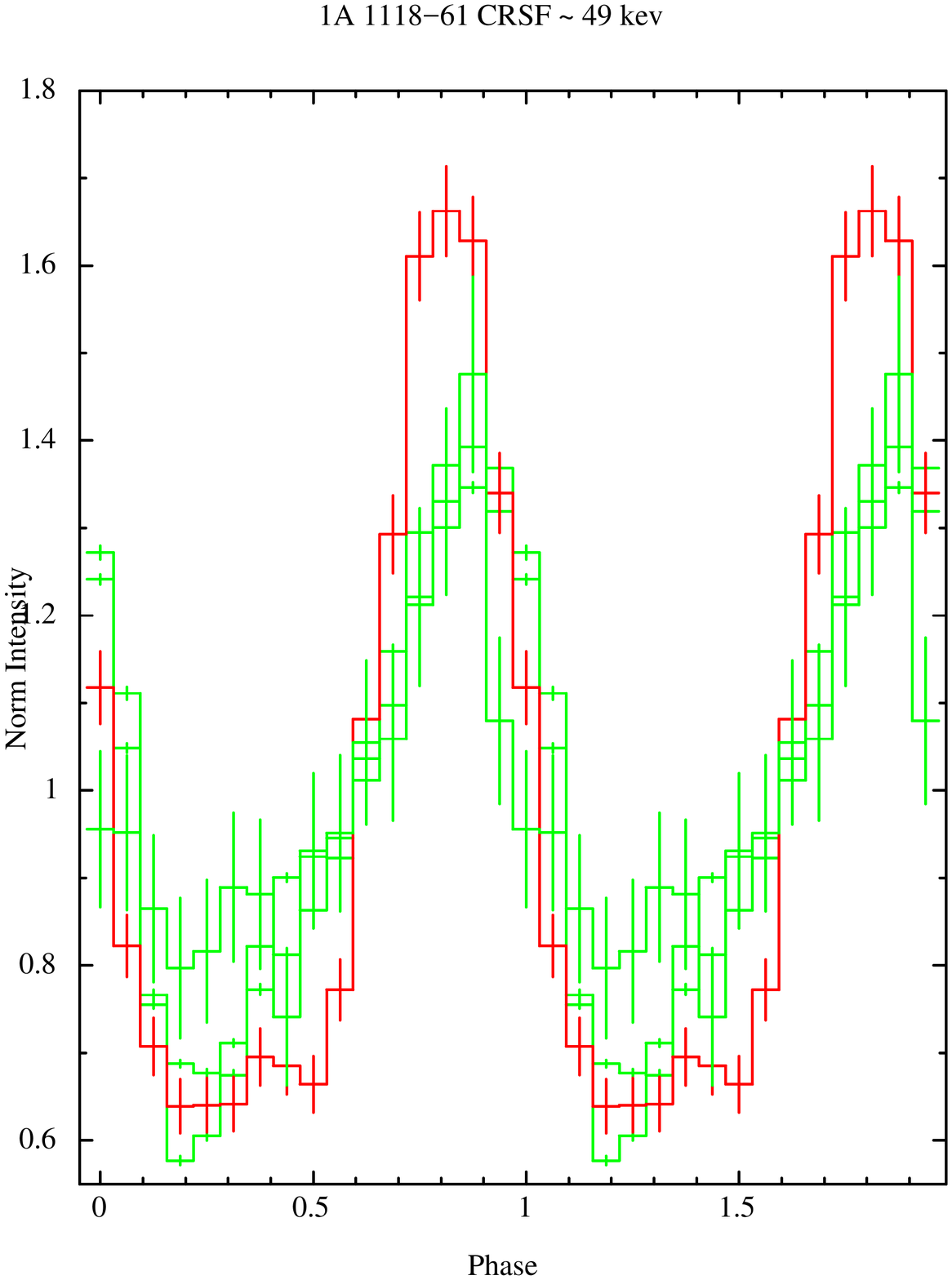}
\includegraphics[scale=0.25]{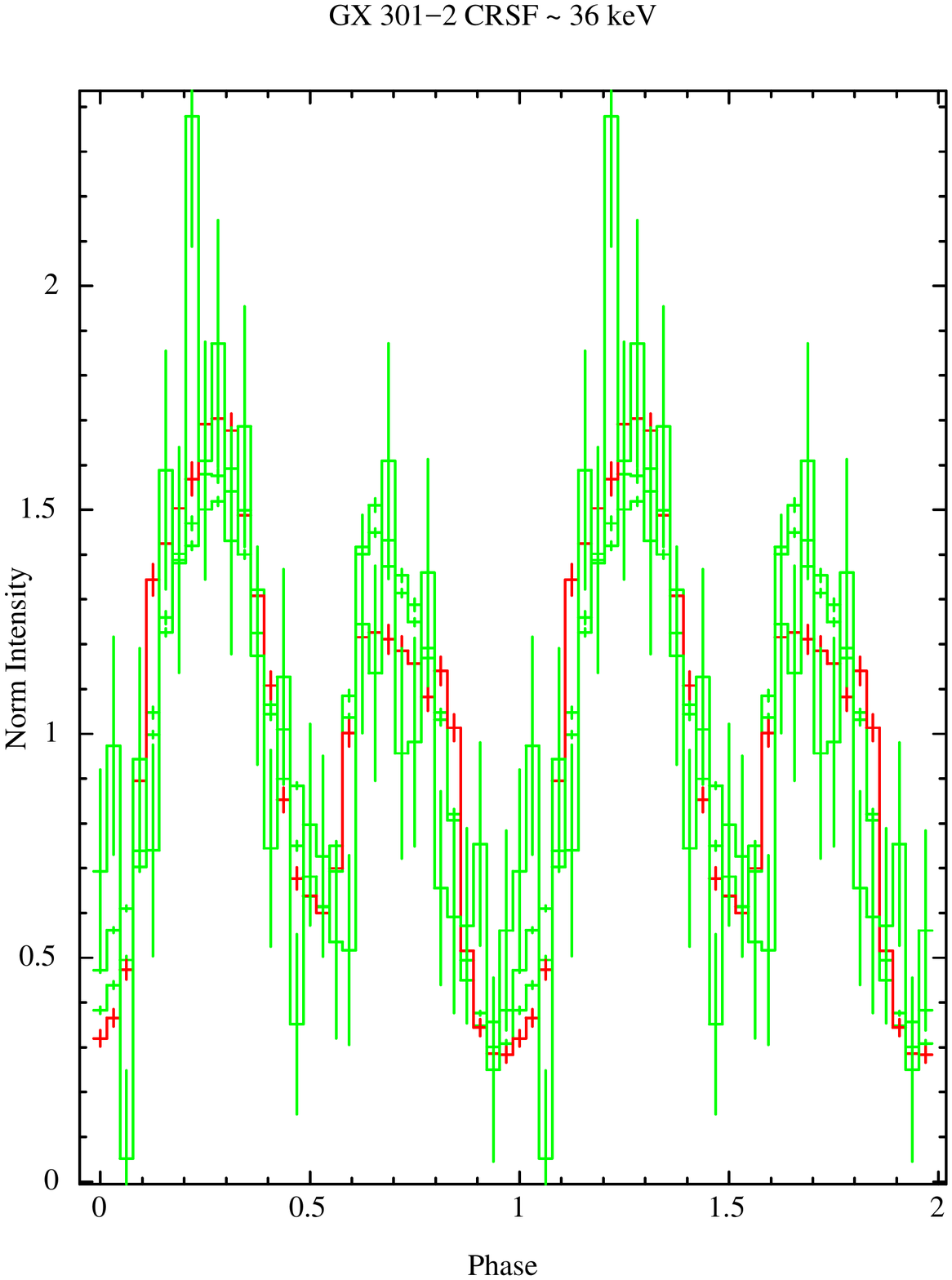}
\includegraphics[scale=0.25]{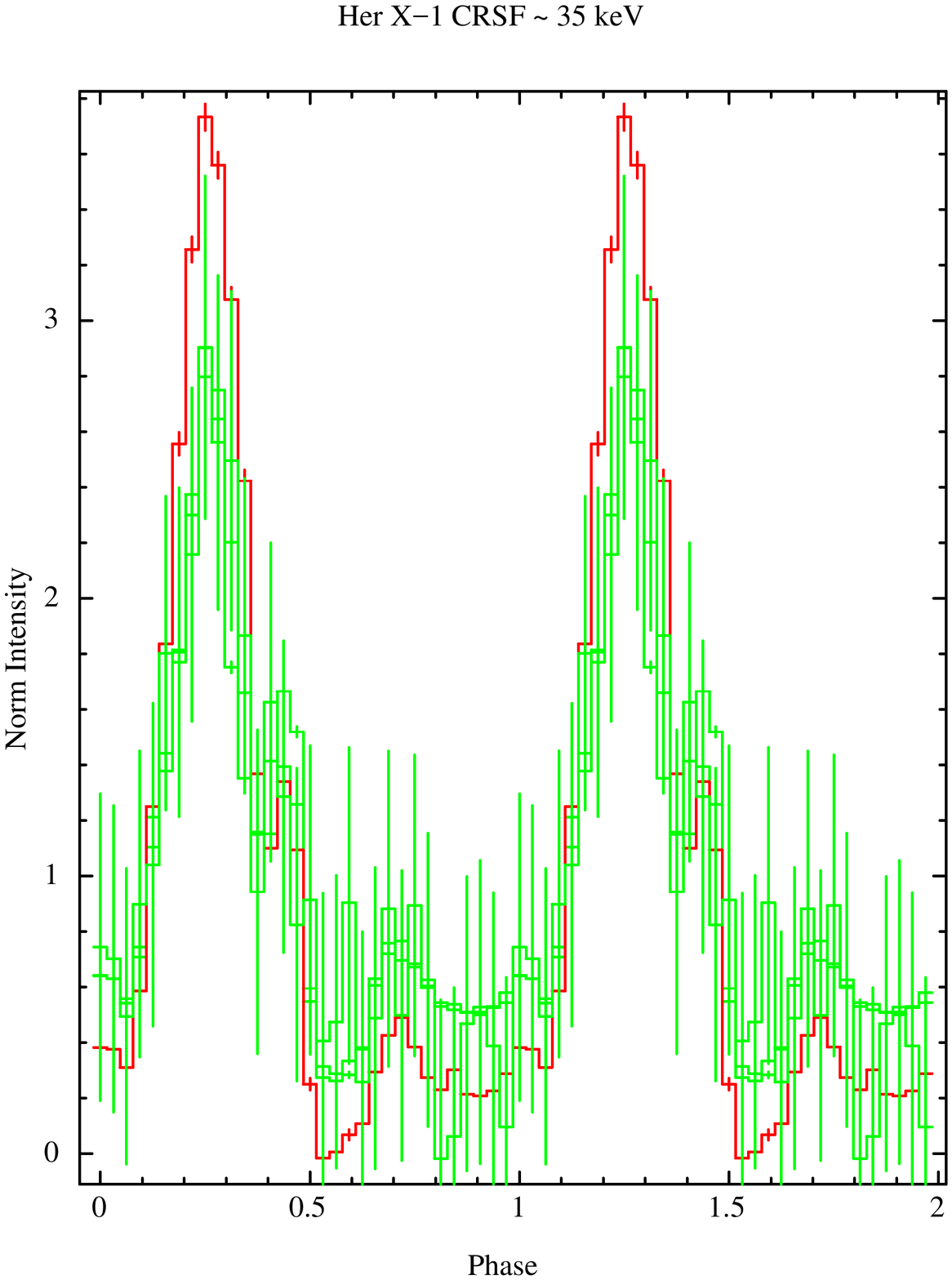}
\caption[Pulse profiles of 1A 1118-61 near the CRSF band]{Energy dependent pulse profiles of 1A 1118-61, GX 301-2 and Her X-1. For all the sources, the pulse profile
near the CRSF band is marked in red, and the others in green (corresponding CRSF energies are marked in figure for respective sources). For all the sources the pulse profiles denote normalized intensity.
( Maitra, C Ph.D Thesis, 2013 }
\label{fig211}
\end{figure*}

 The previous section indicate that timing (studying the pulse shapes in different energy bands) and spectral studies (studying the pattern of change of CRSF with pulse phase) can both provide essential information on the
 line forming region and the accretion geometry in ACPs. Such joint investigations of studying the pulse profile and spectrum of a source at different rotational phases provide an unique opportunity to delve deep into
 the physics of accretion in these systems and map the geometry of the accreting region. To begin with,
  correlation of the pulse phase corresponding to the deepest (and narrowest) and shallowest (and widest) CRSF detected in the spectrum,
   with the pulse profile can indicate the beaming pattern of the source at a given luminosity (a.k.a. accretion rate). A very preliminary demonstration in this direction can be shown with the bright CRSF source Vela X-1 which has relatively well studied timing and spectral properties, with two prominent CRSFs detected at $\sim$ 25 and 50 keV. The pulse profiles (seen in the second panel of Fig.  \ref{vela}) show a clear double peaked profile with very little energy dependence, making it an ideal case study.
  
\begin{figure*}
	\centering
        \includegraphics[width = 8cm, height = 8cm,keepaspectratio] {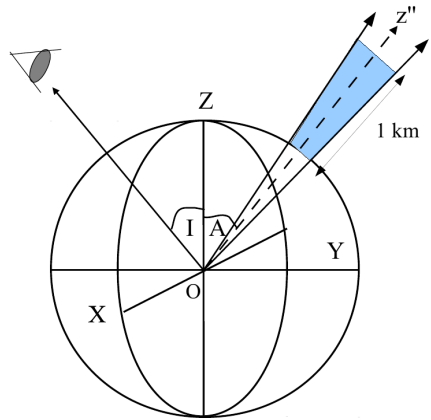}
	\caption{ \small Representative figure showing the geometry of the accretion column for which pulse profiles corresponding to the `Fan' and `Pencil' beam are calculated.}\label{column}
\end{figure*}

  From the results of pulse phase resolved spectroscopy of the CRSF (Fig. \ref{vela}), distinct `Fan' and `Pencil' beam patterns can be identified from trends in variation of the CRSF fundamental. The two peaks of the pulse profile are identified with the trend expected from a `Fan' beam emission (with the deeper and narrowest CRSFs),  and the minima with a `Pencil' beam emission (phases $\sim$ 0.5 - 0.6, with shallower and wider CRSFs). To construct a simple geometrical model corresponding to this,
 pulse profiles are simulated considering the accretion column as a part of a cone whose apex is at the centre of the NS as in Fig. \ref{column} (see \cite{riffert1988} and \cite{leahy2003}). The cone is chosen with an opening angle $\sim$ 5.7$^{\circ}$ so that at r $\sim$ 10 km, the distance from the axis is $\sim$ 1 km, the canonical size of the polar cap radius. The angle $A$ is the inclination angle of the magnetic axis and the angle $I$ is the inclination of the line of sight, both measured with respect to the spin axis ($Z$ in Fig. \ref{column}). The height of the column is taken as $\sim$ 1 km. Emission from the walls of the column are considered to constitute the `Fan' beam, and from the bottom to constitute the `Pencil' beam. Only isotropic emission are considered. The effect of light-bending due to gravity and gravitational redshift of the intensity are considered \citep{beloborodov2002}. The effects of solid angle transformation between the object and image plane are evaluated by a method similar to \cite{2003MNRAS.343.1301P}. The pulse profiles are created for $I=80^\circ$. Vela X-1 being an eclipsing binary, the assumed $I$ is reasonable. For simplicity $A$ is assumed equal to $I$. Due to the high value of $I$, emission from both the magnetic poles
 of the NS are expected to contribute to the pulse profile. The observed pulse profile is reconstructed by adding the `Fan' and `Pencil' emission components from each pole so that we obtained a qualitative match with the observed profile. The ratio of intensities obtained are in Table \ref{tableres}.
 
\begin{figure*}
	\centering
       \includegraphics[width = 10cm, height = 10cm,keepaspectratio] {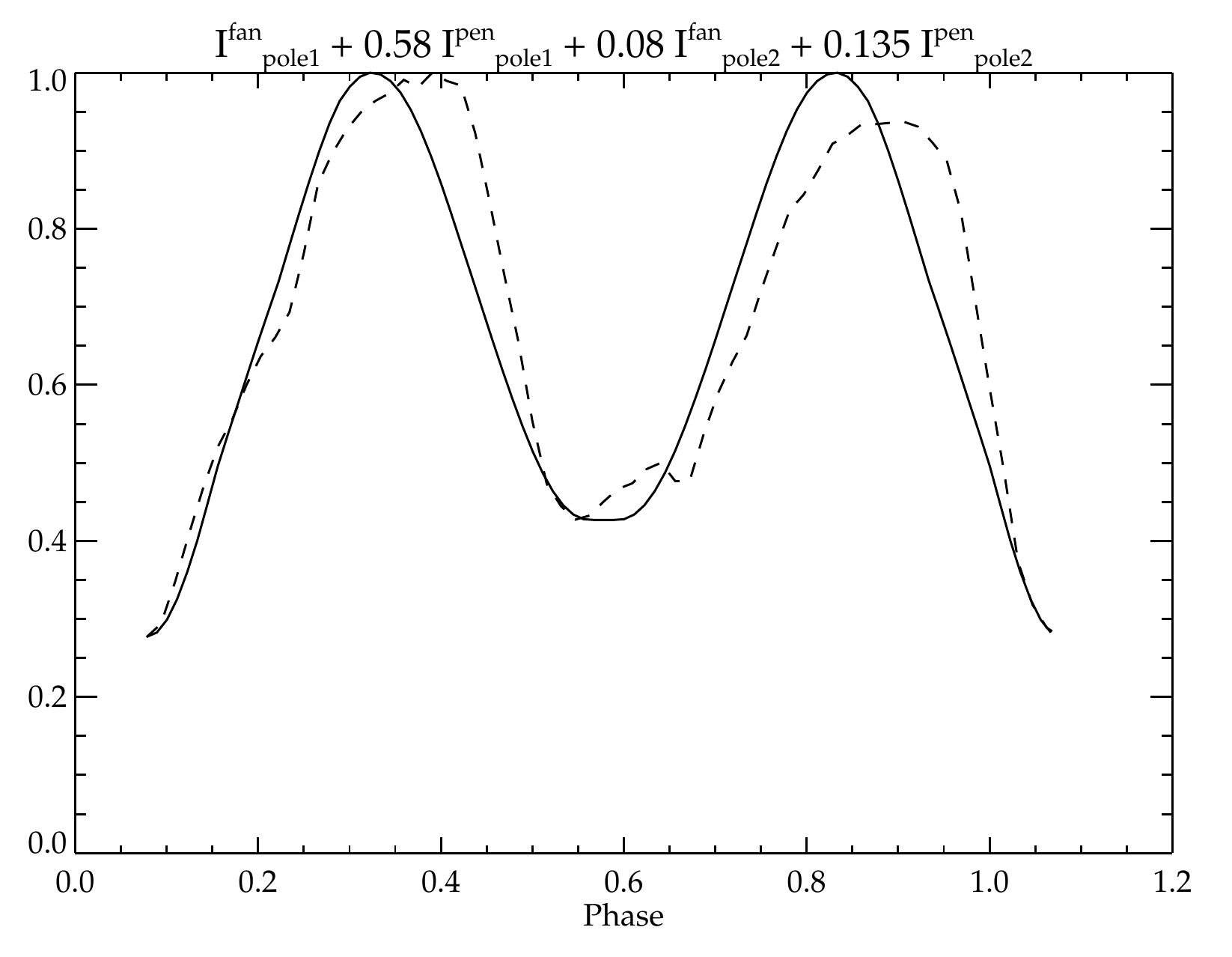}
        \includegraphics[width = 10cm, height = 10cm,keepaspectratio] {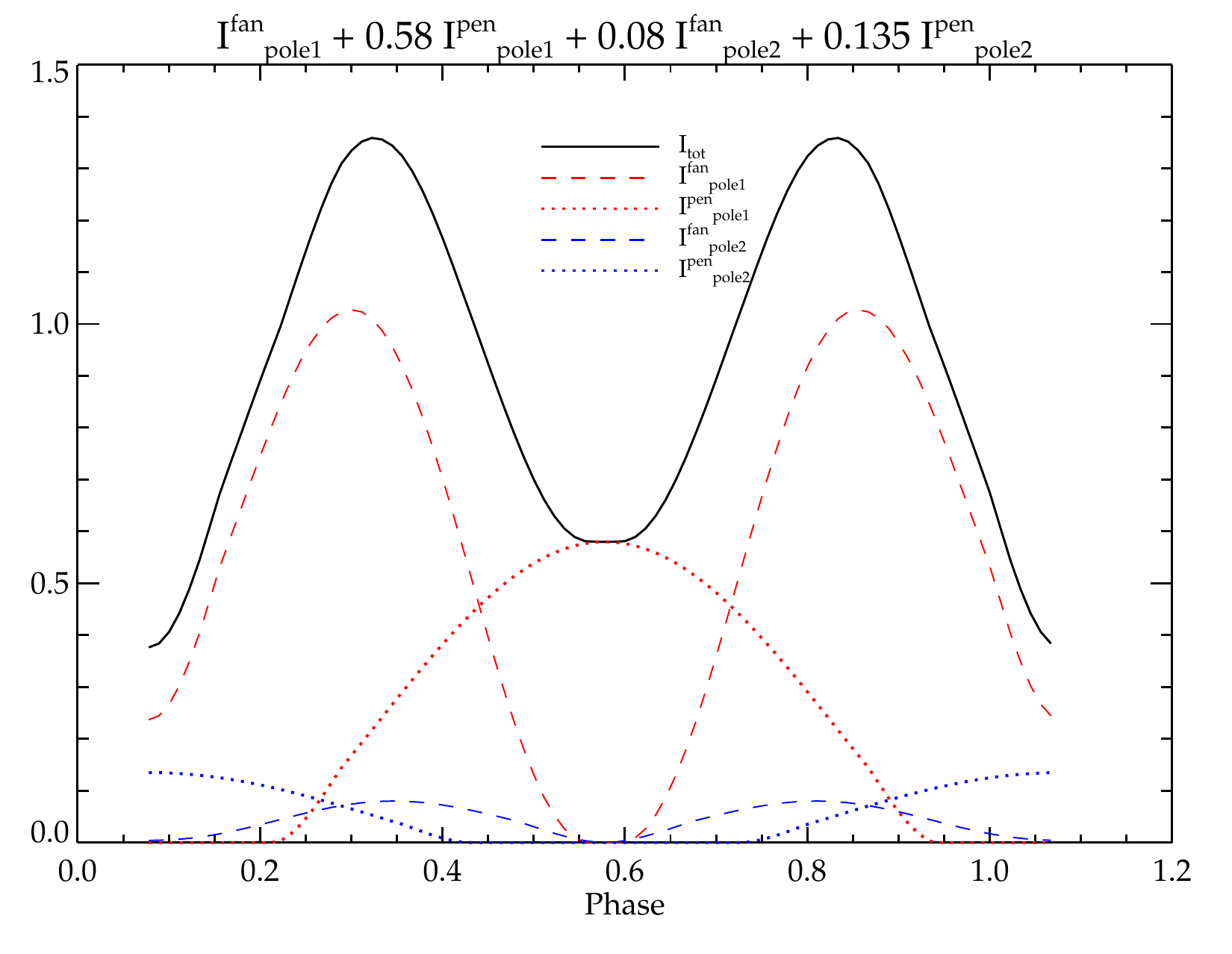}
	\caption{ \small Top: Comparison of the simulated pulse profile of Vela X-1 (dashed curve) with the observed (solid curve). Bottom: Decomposition of the simulated pulse profile of Vela X-1 showing contribution from the different components from both the magnetic poles marked in the figure. The nearest magnetic pole is marked with designation 1.  $A=80^\circ$ and {\it l.o.s} at $I=80^\circ$.}\label{sim-lc}
\end{figure*}

 Although the simple model cannot exactly reproduce the shape of the profile, the simulated profile matches fairly well as is seen in Fig. \ref{sim-lc}. A phase offset is seen between the peaks of the pulse profile, and peak of `Fan' beams which can be due to an asymmetric beaming (this can result into offsets between peaks of a `Fan' beam). The observed pulse peaks are also asymmetric (gradual rise and sharp fall) which indicates the requirement to take into effect more factors like a complex column geometry or asymmetric accretion onto the poles. The angles ($A$ and $I$) also need to be refined and the next step is to fit the simulated profile with the observed one with  $A$ , $I$, $I_{\rm Fan}$ and $I_{\rm Pencil}$ as free parameters.
 Fitting of more complicated profiles would also require an energy dependence to be incorporated.  The obtained results can then be used to investigate whether the observed CRSF trend for a given source can also be reproduced simultaneously by adding simulated CRSF spectra corresponding to the `Fan' and `Pencil' components, in the same ratio indicated by the pulse profile fit (Mukherjee et al., in preparation)
 \begin{table}
\centering
\small
\begin{tabular}{|l|l|}
\hline
Intensity components & Fraction \\
\hline
 1st pole(nearest) & \\
$I_{\rm Fan}$ & 1.0 \\
$I_{\rm Pencil}$ & 0.58 \\
\hline
2nd pole & \\
$I_{\rm Fan}$ & 0.08 \\
$I_{\rm Pencil}$ & 0.135 \\
\hline
\end{tabular}
\caption{Intensity contribution of the beam patterns for the simulated pulse profile of Vela X-1 from both the magnetic poles.}
\label{tableres}
\end{table}





\section{Conclusion}
The last decades have provided plenty of surprises related to the study CRSFs with complex dependencies and trends of the lines with various physical parameters being revealed.
Changes of the line parameters with luminosity have provided a probe into the change of accretion geometry with variations in the accretion rate. Mapping the CRSF at different
viewing angles (and hence the rotational phase of the NS) have provided insights into the distribution of the magnetic field, plasma temperature and optical depth in the line forming region,
as well as the beaming pattern and the accretion geometry itself.  Detection of complex shapes of CRSFs have provided clues into what might be an effect due magnetic field distortions and instabilities.
A trend in the long term evolution in the energy of the line has indicated a secular evolution of the NS magnetic field itself. Significant efforts have also been put to model 
these features taking into account as many physical processes as allowed by computational limitations across the parameter space relevant to these systems. The future aims at more sensitive studies in the hard X-ray band ($>$ 10 keV) where CRSFs are primarily detected in accreting neutron stars. The ongoing missions \nustar~ and~{\it ASTROSAT} are ideal in this regard. Modelling
the timing and spectral results jointly with the latest physical models can provide a comprehensive picture on the physics of these systems.

\section*{Acknowledgements}
Many results discussed in this review regarding the pulse phase resolved spectroscopy of CRSFs and studying the pulse profiles near the CRSF band were obtained as a part of Ph.D thesis of CM under the 
supervision of Biswajit Paul. CM acknowledges very insightful discussions and suggestions from Dipankar Bhattacharya which helped her understand the physics of CRSFs better. The geometrical model discussed in section 4
was done in collaboration with Dipanjan Mukherjee and Dipankar Bhattacharya and is work under progress.


\bibliographystyle{mnras}
\bibliography{mnrasmnemonic,crsf2017}

\end{document}